\documentclass[namedreferences]{SolarPhysics}
\usepackage[optionalrh]{spr-sola-addons} % For Solar Physics
\usepackage{graphicx}        % For eps figures, newer & more powerfull
\usepackage{color}           % For color text: \color command
\usepackage{url}             % For breaking URLs easily trough lines
\newcommand{\aap}{    {\it Astron. Astrophys.}}

\newcommand{\apj}{    {\it Astrophys. J.}}

\newcommand{\solphys}{{\it Solar Phys.}}

\newcommand{\ssr}{    {\it Space Sci. Rev.}}

\begin{document}

\begin{article}

\begin{opening}

\title{Investigation of quasi-periodic variations in hard X-rays
 of solar flares. II. Further investigation of oscillating magnetic traps}

%%%%%%%%%%%%%%%%%%%%%%%%%%%%%%%%%%%%%%%%%%%%%%%%%%%
%% Authors Names
%
\author{J.~\surname{Jakimiec}\sep
        M.~\surname{Tomczak}%$^{1}$\sep
%        I.~\surname{}$^{2}$
       }

%%%%%%%%%%%%%%%%%%%%%%%%%%%%%%%%%%%%%%%%%%%%%%%%%%%
%% Runningheads
%
\runningauthor{J.\,Jakimiec \& M.\,Tomczak} \runningtitle{QPO in
HXRs of solar flares. II}

%%%%%%%%%%%%%%%%%%%%%%%%%%%%%%%%%%%%%%%%%%%%%%%%%%%
%% Affilations
%
  \institute{Astronomical Institute, University of Wroc{\l }aw,
  ul. Kopernika 11, 51-622 Wroc{\l }aw, Poland,
                     email: \url{jjakim; tomczak@astro.uni.wroc.pl}\\
%             $^{2}$ Second affiliation
%                     email: \url{e.mail-c} \\
             }

\begin{abstract}
In our recent paper (Solar Physics {\bf 261}, 233) we investigated
quasi-periodic oscillations of hard X-rays during impulsive phase of
solar flares. We have come to conclusion that they are caused by
magnetosonic oscillations of magnetic traps within the volume of
hard-X-ray (HXR) loop-top sources. In the present paper we
investigate four flares which show clear quasi-periodic sequences of HXR pulses. We also describe our phenomenological model of oscillating magnetic traps to show that it can explain observed properties of HXR oscillations. Main results are the following:
\begin{enumerate}
\item We have found that low-amplitude quasi-periodic oscillations occur before impulsive phase of some flares.
\item We have found that quasi-period of the oscillations can change in some flares. We interpret this as being due to changes of the length of oscillating magnetic traps.
\item During impulsive phase a significant part of the energy of accelerated (non-thermal) electrons is deposited within the HXR loop-top source.
\item Our analysis suggests that quick development of impulsive phase is due to feedback between pulses of the pressure of accelerated electrons and the amplitude of magnetic-trap oscillation.
\item We have also determined electron number density and magnetic filed strength for HXR loop-top sources of several flares. The values fall within the limits of $N \approx (2 -15) \times 10^{10}$ cm$^{-3}$, $B \approx (45 - 130)$ gauss.
\end{enumerate}
\end{abstract}
%%%%%%%%%%%%%%%%%%%%%%%%%%%%%%%%%%%%%%%%%%%%%%%%%%%
%% Keywords
%

\end{opening}
%-------------------------------------------------

\section{Introduction}
     \label{intr}

In hard X-ray (HXR) emission of many flares quasi-periodic variations were observed with time-intervals between pulses, $P \sim 10-60$\,s [see \inlinecite{lip78}; see also review of \inlinecite{n+m09} and references therein].

In our previous paper (\opencite{paper1}, Paper I) we attempted to investigate relationship between HXR loop-top (LT) sources and the quasi-periodic variations. Main difficulty was that sequences of pulses are usually short, so that it is difficult to carry out comprehensive analysis of their quasi-periodicity. Therefore we used time-interval, $\Delta{t}$, between the strongest pulses as a simple estimate of the quasi-period $P$.

In the present paper we have selected four flares which have longer sequences of HXR pulses, so that it was possible to carry out detailed analysis of their quasi-periodicity (Section \ref{qper}). Section \ref{16jan} contains detailed analysis of 16 January 1994 flare. In Section \ref{det} we estimate values of electron density and magnetic field strength inside several HXR loop-top sources. Section \ref{sum} contains discussion and conclusions.

\section{Observations and their analysis}\label{obs}

We used HXR observations recorded by {\sl Yohkoh} Hard X-ray Telescope, HXT, \cite{kos91} [light-curves and images] and {\sl Compton Gamma Ray Observatory} Burst and Transient Source Experiment, BATSE, \cite{fis92} [light-curves].

\subsection{Analysis of quasi-periodicity of HXR pulses}\label{qper}

\begin{figure}
\centerline{\includegraphics[width=1\textwidth]{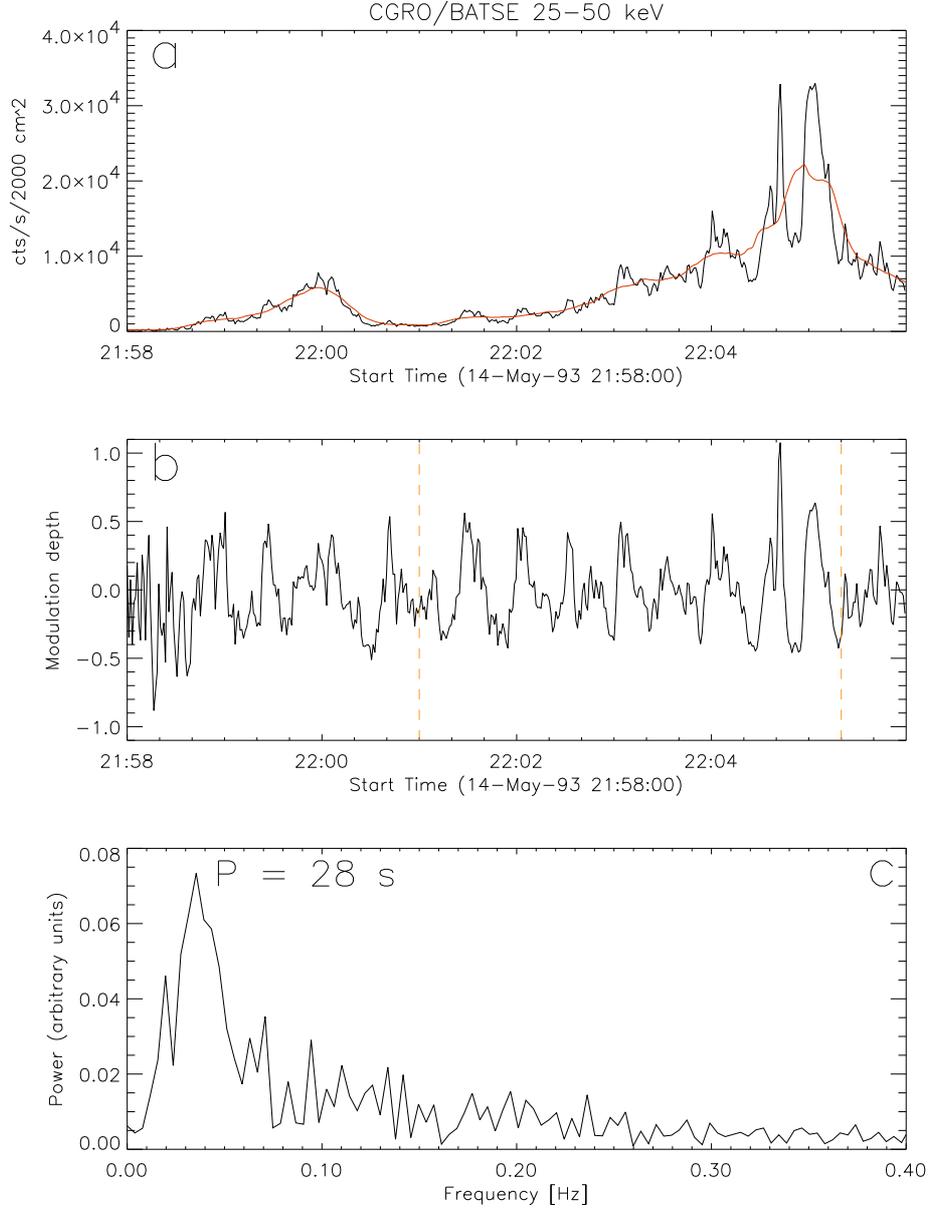}}
\hfill \caption{Analysis of the HXR light-curve (25-50 keV, {\sl Compton Gamma Ray Observatory}/BATSE observations) for a flare of 14 May 1993. (a) HXR light-curve. The red line shows running average calculated with averaging time ${\delta}t = 30$ s. (b) Normalized light-curve, $S(t)$ [see Equation (1)]. Vertical dashed lines show the time-interval which was used to calculate power spectrum. (c) Power spectrum calculated for the normalized light-curve, $S(t)$. $P$ is the period corresponding to the peak in power spectrum.} \label{93may14pgm}
\end{figure}

We have selected four flares which have longer sequences of HXR pulses, so that detailed analysis of their quasi-periodicity was possible. Main difficulty in the analysis was the fact that during impulsive phase the pulses occur simultaneously with quick increase of total HXR intensity (see Figure \ref{93may14pgm}a). Therefore we have applied a method of normalization which is commonly used by radioastronomers (see \opencite{fle08}). A normalized time series, $S(t)$, is:
\begin{equation}
S(t) = \frac{F(t) - \hat{F}(t)}{\hat{F}(t)},
\end{equation}
where $F(t)$ is the measured HXR flux and $\hat{F}(t)$ is a running average of $F(t)$. The red line in Figure \ref{93may14pgm}a shows $\hat{F}(t)$ calculated with averaging time $\delta{t} = 30$\,s. The normalized time series, $S(t)$ is shown in Figure \ref{93may14pgm}b. Quasi-periodicity of the pulses is clearly shown in this Figure and it is confirmed by power spectrum seen in Figure \ref{93may14pgm}c (Fourier transform of $S(t)$). The power spectrum has been calculated for time-interval 22:01-22:05:20 UT (Impulsive Phase, IP).

We see in Figures \ref{93may14pgm}a and \ref{93may14pgm}b that before the impulsive phase, between 21:58:40 and 22:00:20 UT, three increasing pulses occurred. Mean time-interval between the pulses is about 31\,s which is close to the quasi-period, $P$, seen in the power spectrum (28\,s). This suggests that these pulses also belong to the quasi-periodic sequence seen during impulsive phase.

\begin{figure}
\centerline{\includegraphics[width=1\textwidth]{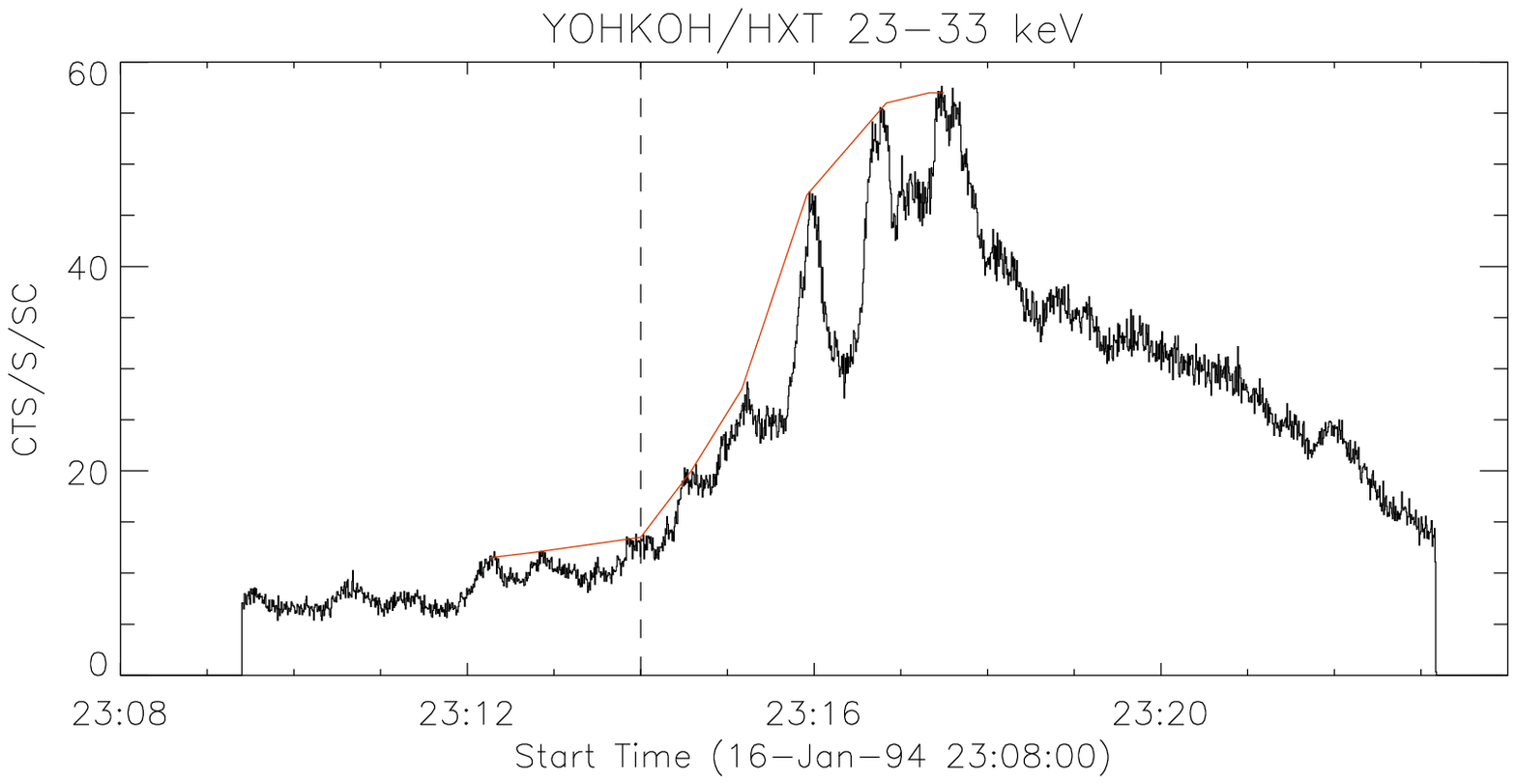}}
\hfill \caption{23-33 keV light-curve recorded by {\sl Yohkoh}/ Hard X-ray Telescope for a flare of 16 January 1994. The red line connects peaks of HXR pulses to visualize their non-linear increase and saturation. The noise is due to statistical fluctuations of counting rate. The dashed vertical line shows the beginning of impulsive phase.} \label{94jan16lc}
\end{figure}

\begin{figure}
\centerline{\includegraphics[width=1\textwidth]{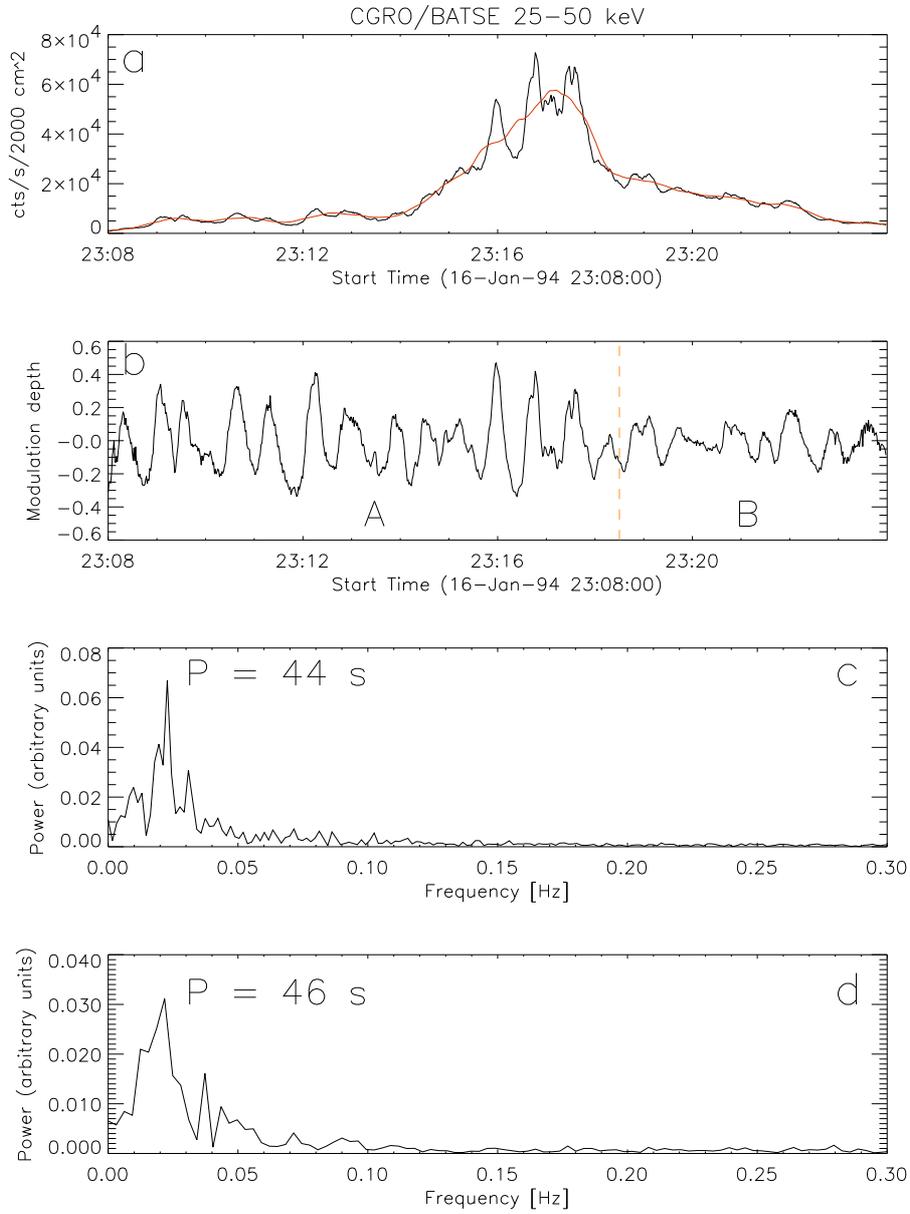}}
\hfill \caption{The same as in Figure 1, for a flare of 16 January 1994. the power spectra have been calculated for time intervals A and B and they are shown in panels (c) and (d).} \label{94jan16pgm}
\end{figure}

Figure \ref{94jan16lc} shows a HXR light-curve of 16 January 1994 flare. Like in Figure \ref{93may14pgm}a, during impulsive phase we see a sequence of increasing pulses. The red line connects tops of the pulses to show their quick non-linear increase and saturation. Before the impulsive phase (BIP) there is a sequence of weak pulses. Normalization and power-spectrum analysis is shown in Figure \ref{94jan16pgm}. The power spectrum in Figure \ref{94jan16pgm}c has been calculated for time interval A, i.e. BIP plus IP. Figure \ref{94jan16pgm}d shows power spectrum calculated for the decay phase (time-interval B), where the pulses were weak and their profiles were disturbed, but the quasi-period had been retained.

\begin{figure}
\centerline{\includegraphics[width=1\textwidth]{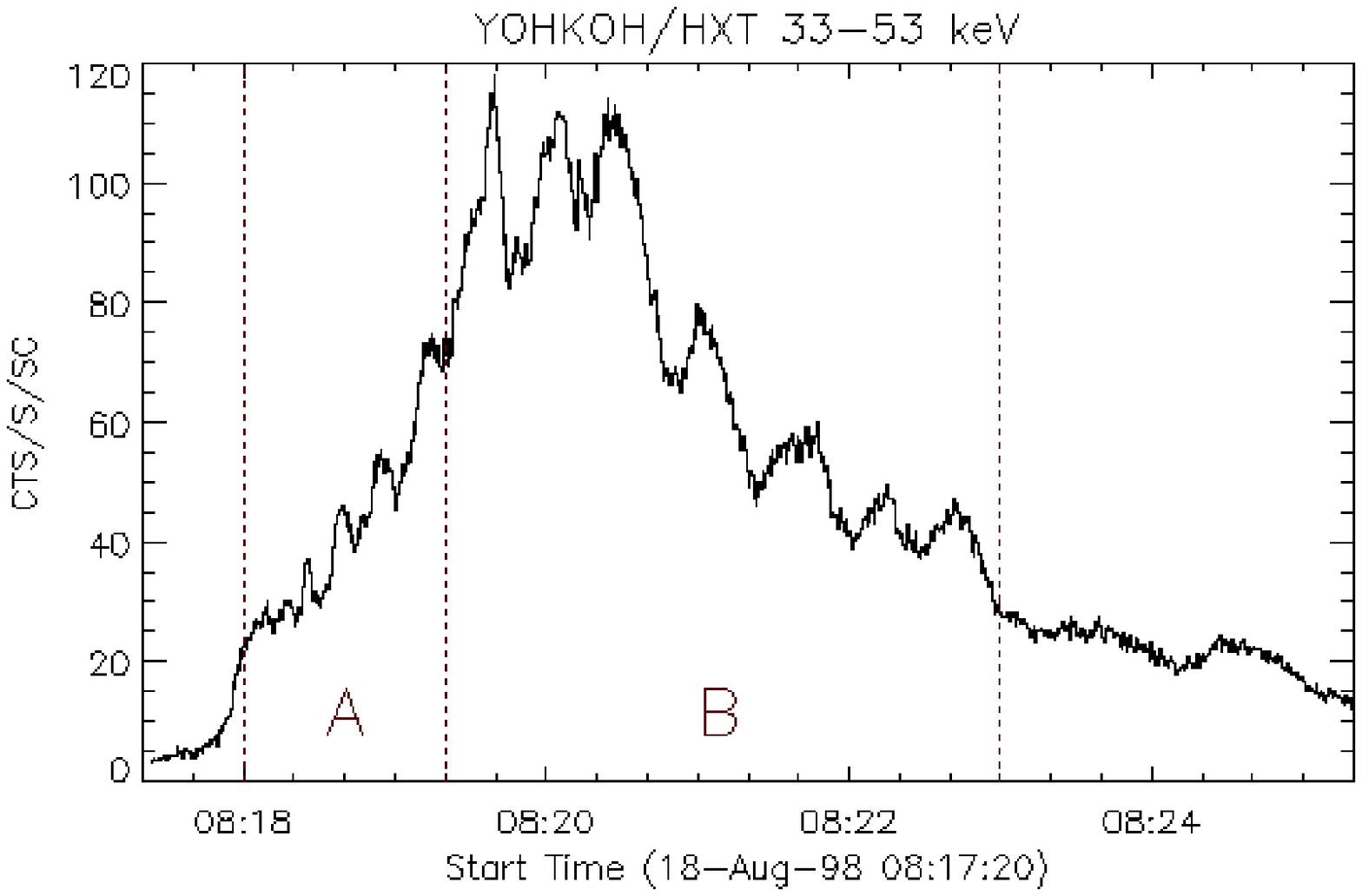}}
\hfill \caption{33-53 keV {\sl Yohkoh}/HXT light-curve for a flare of 18 August 1998. Power spectrum analysis has been carried out for time intervals A and B, separately, and results are presented in Figures 5 and 6.} \label{98aug18lc}
\end{figure}

\begin{figure}
\centerline{\includegraphics[width=1\textwidth]{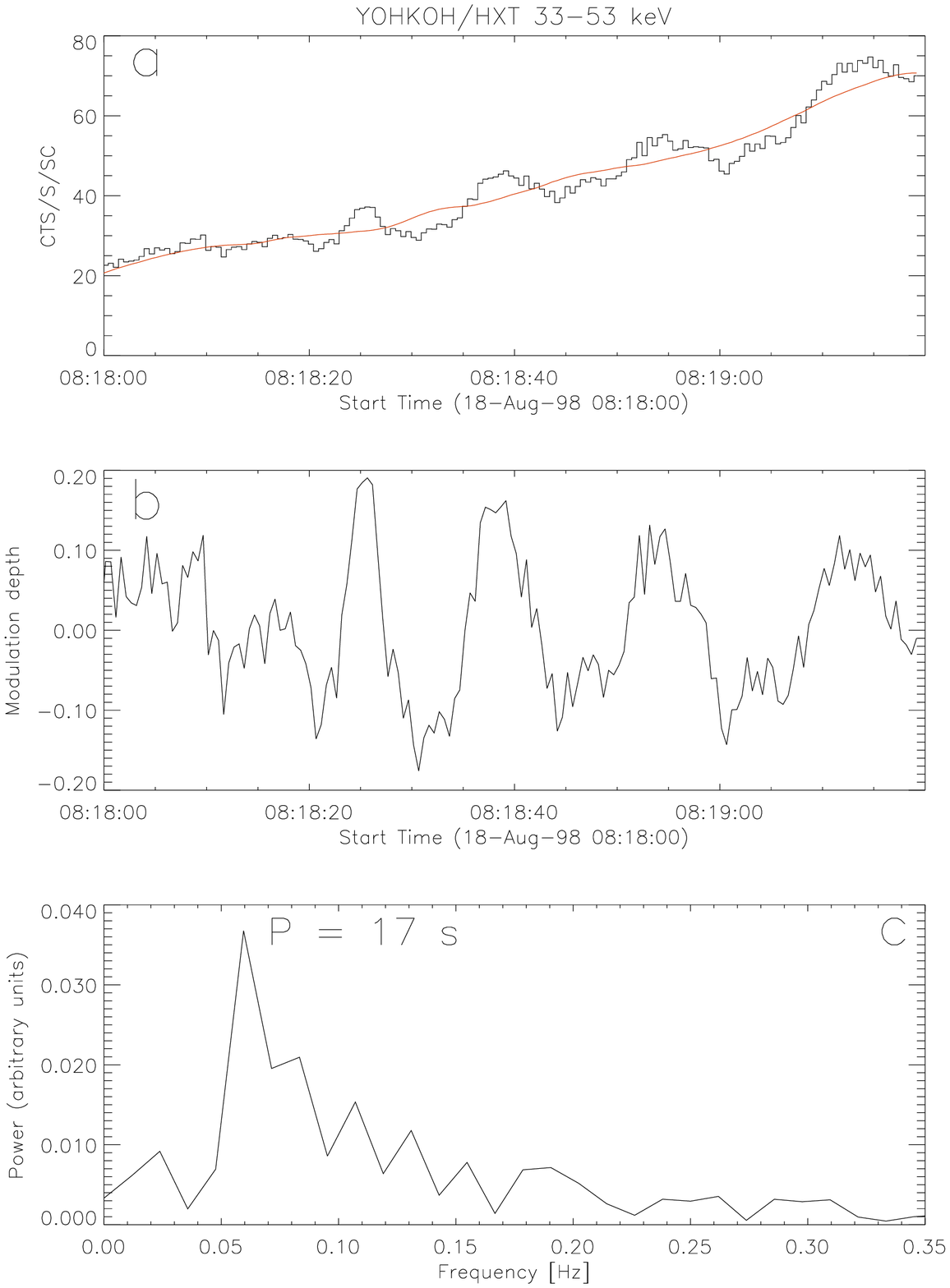}}
\hfill \caption{Power spectrum analysis for time-interval A of the flare of 18 August 1998.} \label{98aug18pgma}
\end{figure}

\begin{figure}
\centerline{\includegraphics[width=1\textwidth]{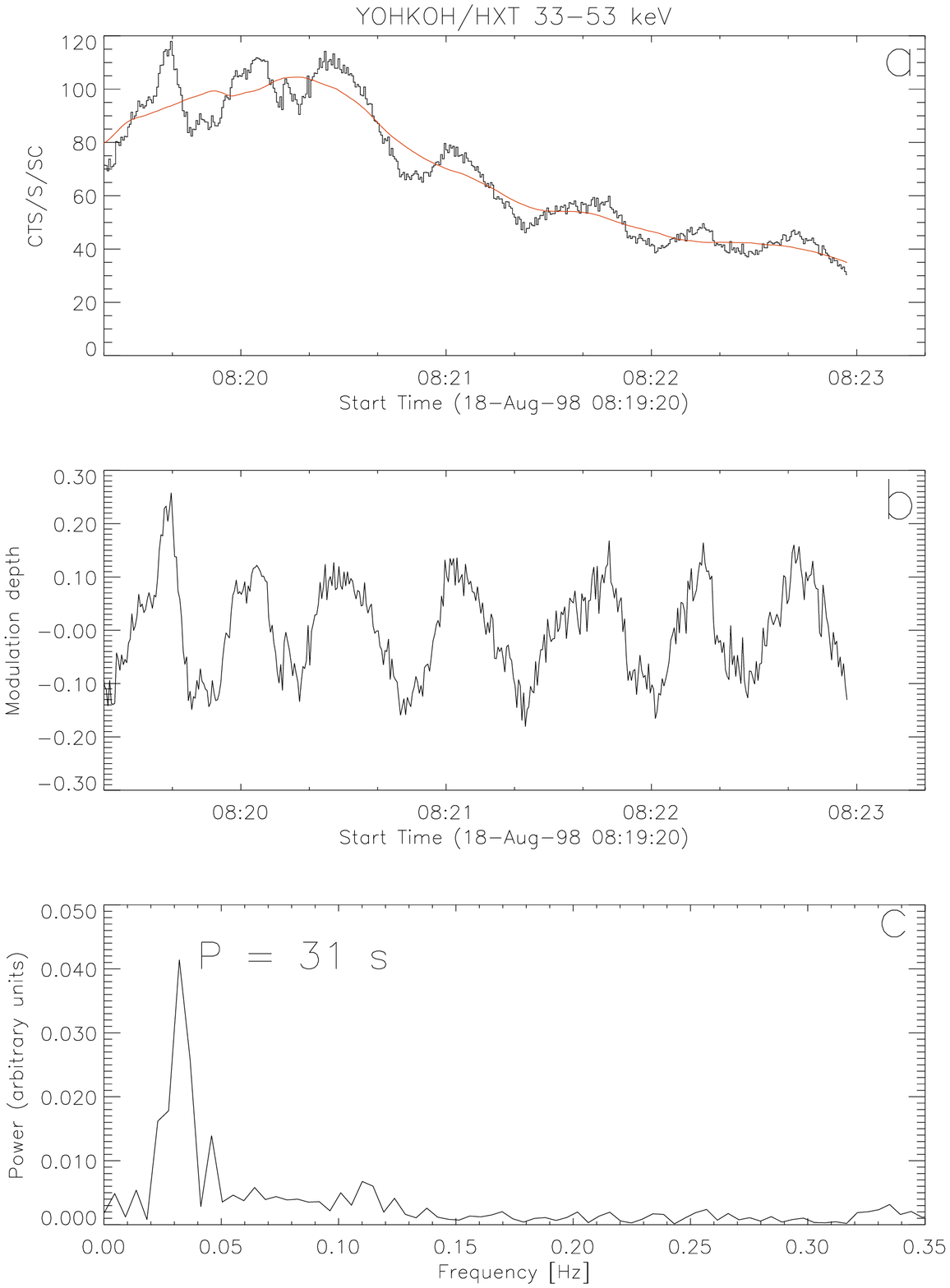}}
\hfill \caption{Power spectrum analysis for time-interval B of the flare of 18 August 1998.} \label{98aug18pgmb}
\end{figure}

Figure \ref{98aug18lc} shows HXR light-curve of 18 August 1998 flare. Its characteristic feature is that quasi-period during increase of HXR emission is shorter than during decrease of the emission. Therefore analysis of the light-curve has been done separately for time-intervals A and B (see Figures \ref{98aug18pgma} and \ref{98aug18pgmb}). In our model of oscillating magnetic traps changes of quasi-period are explained as being due to changes of length of oscillating traps (see Section 3).

\begin{figure}
\centerline{\includegraphics[width=1\textwidth]{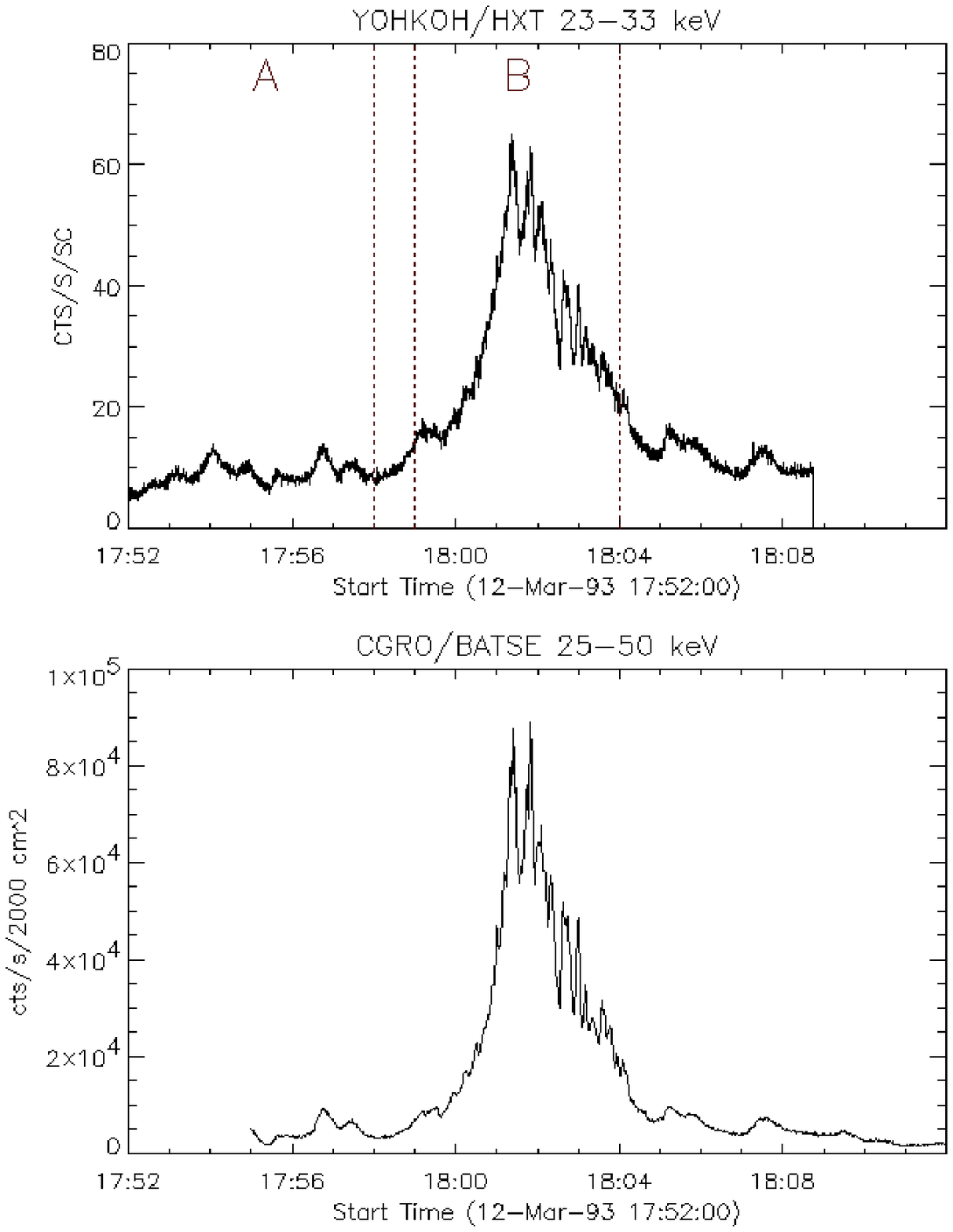}}
\hfill \caption{HXR light-curves for a flare of 12 March 1993 recorded by {\sl Yohkoh}/HXT and {\sl CGRO}/BATSE. Very good agreement of the fluctuations in both curves is seen. Power spectrum analysis has been carried out for time-intervals A and B which are marked by vertical lines.} \label{93mar12lc}
\end{figure}

\begin{figure}
\centerline{\includegraphics[width=1\textwidth]{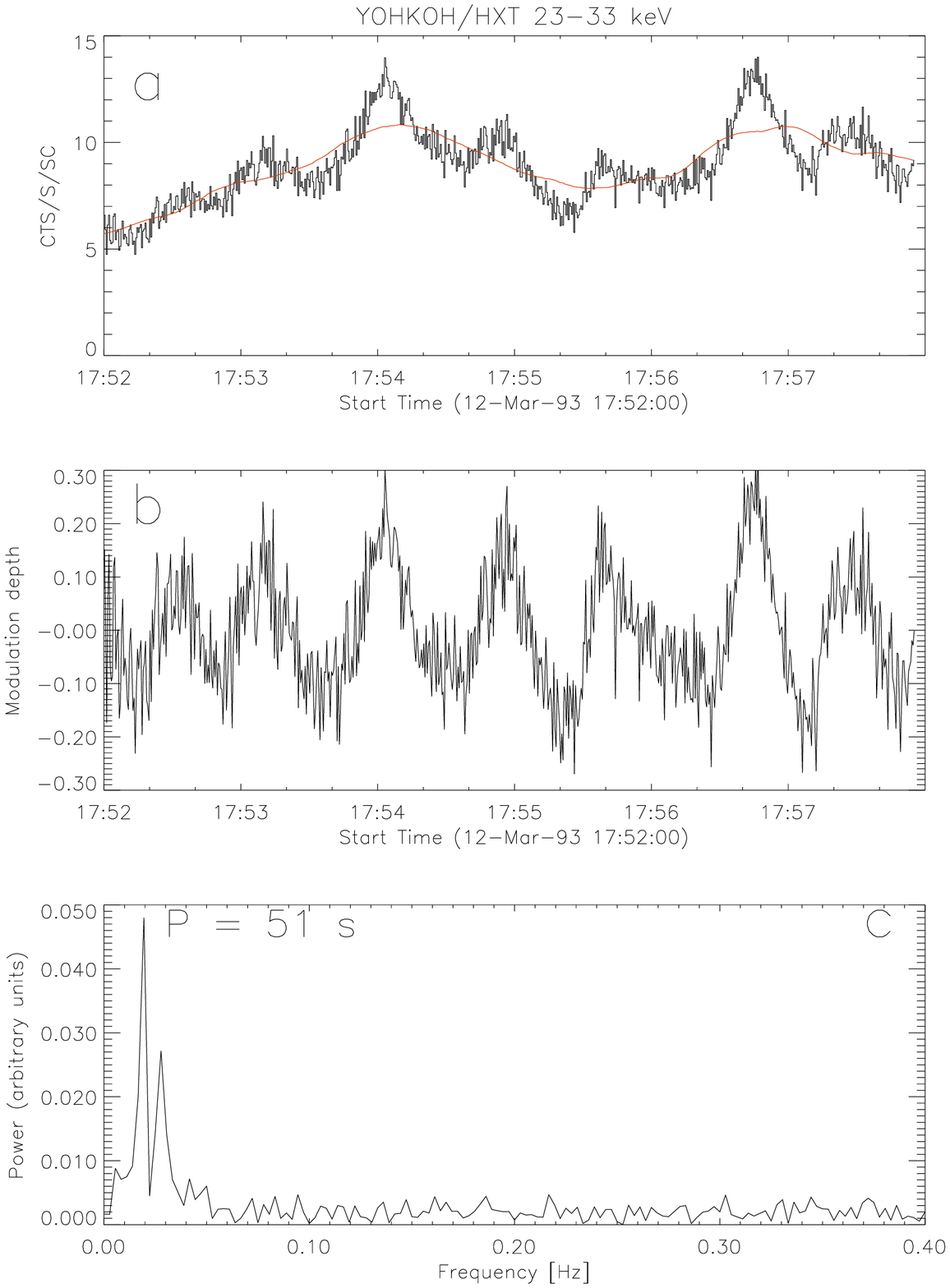}}
\hfill \caption{Power spectrum analysis for time-interval A of the flare of 12 March 1993.} \label{93mar12pgma}
\end{figure}

\begin{figure}
\centerline{\includegraphics[width=1\textwidth]{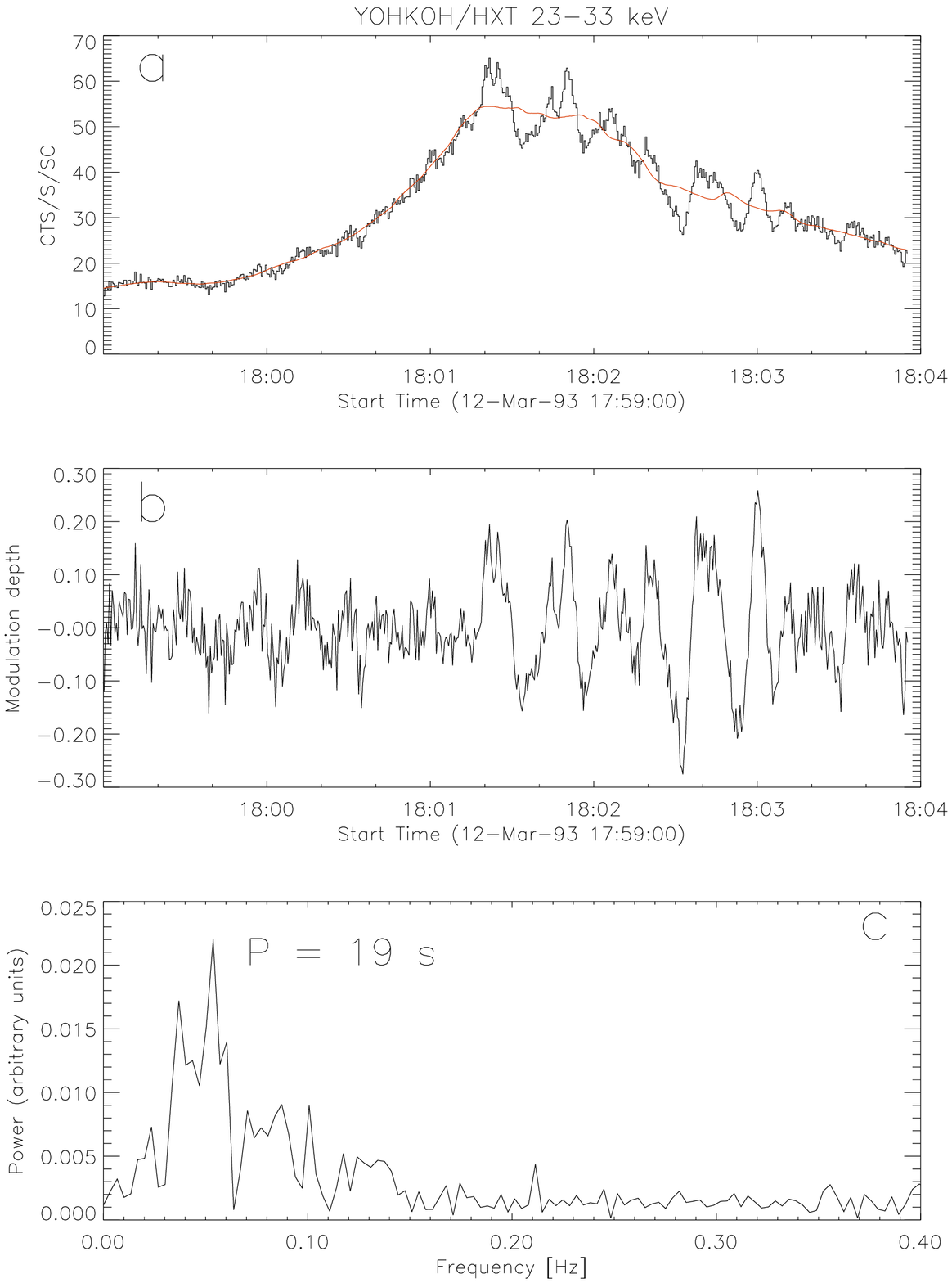}}
\hfill \caption{Power spectrum analysis for time-interval B of the flare of 12 March 1993.} \label{93mar12pgmb}
\end{figure}

Figure \ref{93mar12lc} shows HXR light-curves of 12 March 1993 flare. General behavior is similar to that seen in Figure 2: There are low-amplitude oscillations before impulsive phase and quick increase of HXR emission after the onset of impulsive phase. Specific feature are short-period oscillations after HXR maximum. Therefore analysis of the light-curve has been done separately for time-intervals A and B (see Figures \ref{93mar12pgma} and \ref{93mar12pgmb}). Change of the quasi-period is clearly seen.

\subsection{Analysis of 16 January 1994 flare} \label{16jan}

\begin{figure}
\centerline{\includegraphics[width=1\textwidth]{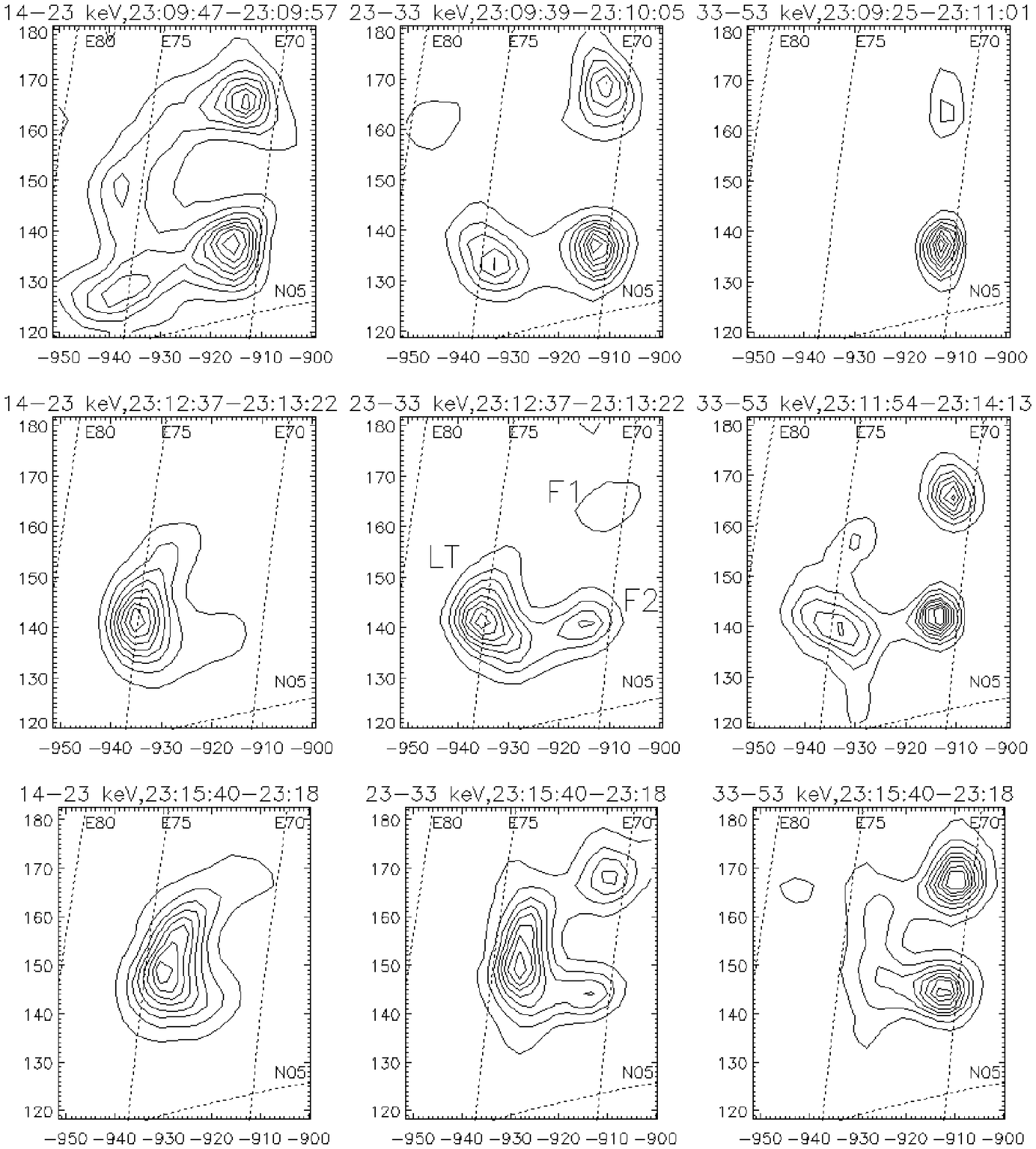}}
\hfill \caption{HXR images of the flare of 16 January 1994 in three
energy channels: 14-23, 23-33, and 33-53 keV (vertical columns).
Upper row: images recorded before impulsive phase, middle row:
recorded near the beginning of impulsive phase, lower row: recorded
during impulsive phase. See text for discussion.} \label{mozaic}
\end{figure}

We have chosen the flare of 16 January 1994 for further analysis because loop-top and footpoint sources can be easily recognized in its HXR images (Figure \ref{mozaic}). We wanted to investigate what happened in the loop-top source during transition from low-amplitude oscillations to quick increase of HXR emission seen in Figures \ref{94jan16lc} and \ref{94jan16pgm}. Toward this end we have reconstructed {\sl Yohkoh}/HXR images in energy channels L (14-23 keV), M1 (23-33 keV), and M2 (33-53 keV) for the time of Before Impulsive Phase (BIP) oscillations (upper row in Figure \ref{mozaic}), about the beginning of impulsive phase (middle row) and for maximum of IP (lower row).

We see that before impulsive phase HXR footpoints are strong which means that accelerated electrons easily escape from the LT source. But at the beginning of impulsive phase the footpoints are weaker than the loop-top source  (F $<$ LT) in 14-23 and 23-33 keV emission which indicates that most of accelerated electrons deposit their energy within the LT source.

\begin{figure}
\centerline{\includegraphics[width=1\textwidth]{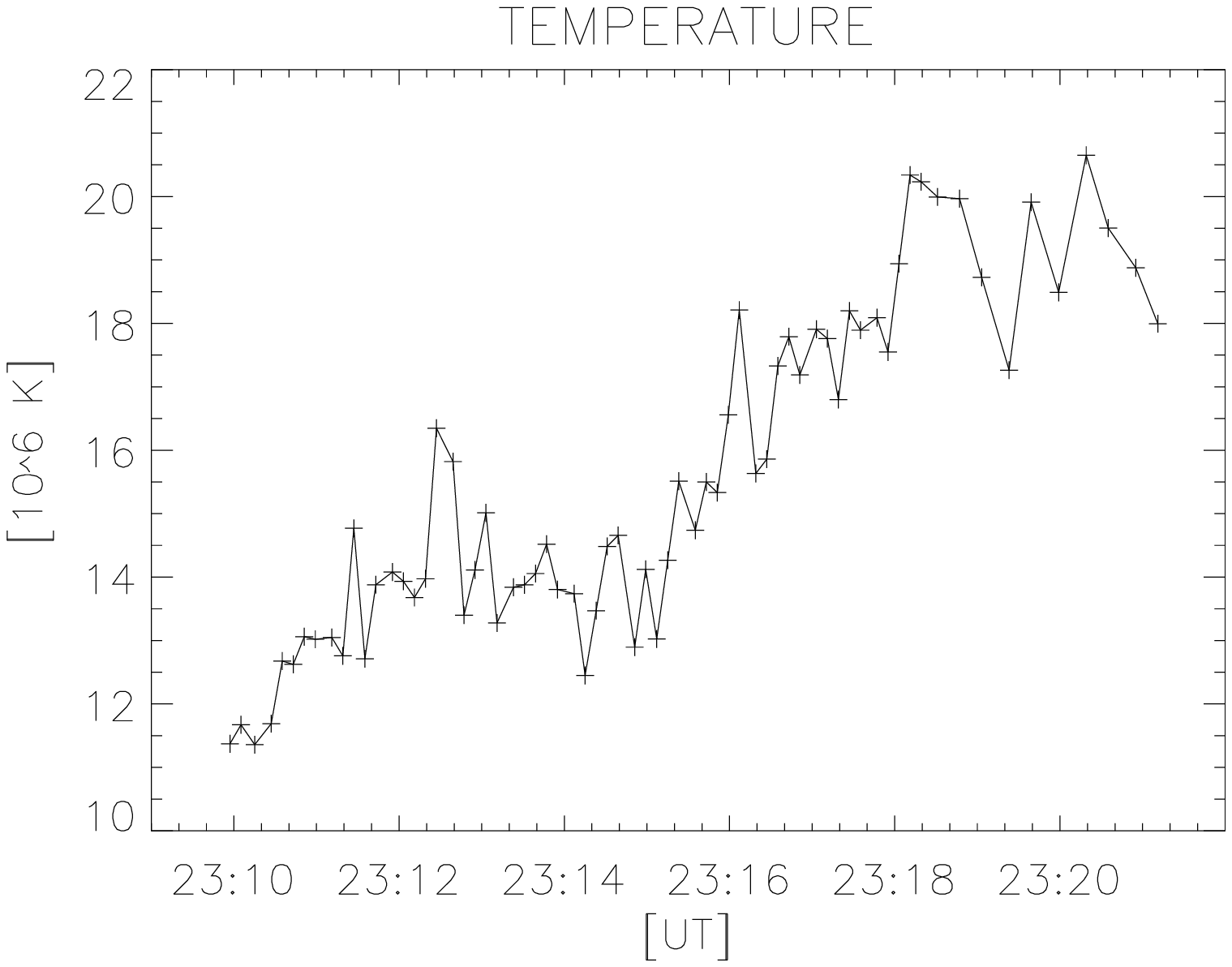}}
\hfill \caption{Time-variation of the mean temperature in HXR
loop-top source of 16 January 1994 flare.} \label{temp}
\end{figure}

\begin{figure}
\centerline{\includegraphics[width=1\textwidth]{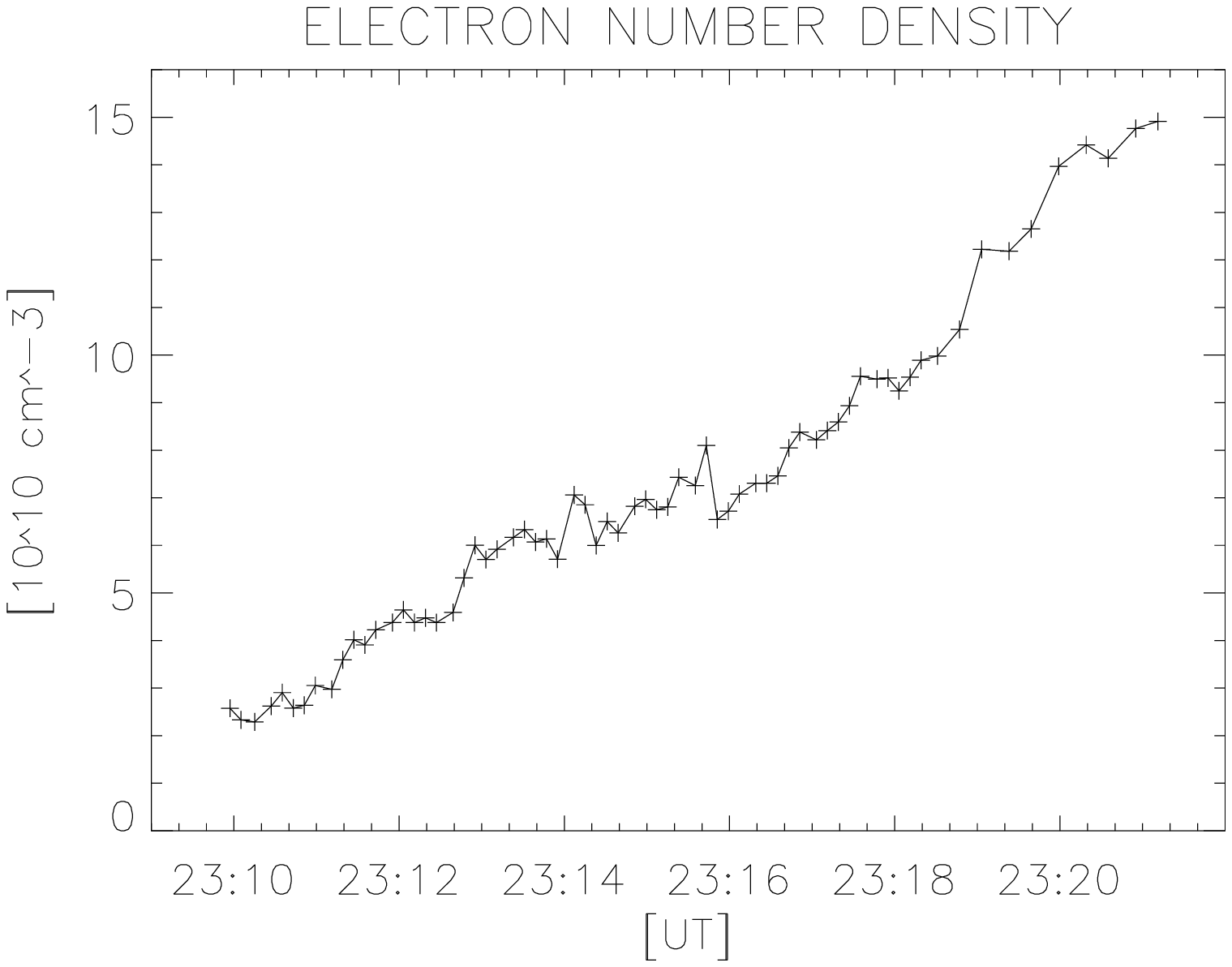}}
\hfill \caption{Time-variation of the mean electron number density
in HXR loop-top source of 16 January 1994 flare.} \label{nel}
\end{figure}

Next we have determined mean temperature, $T$, emission measure,
$EM$, and mean electron number density, $N$, in the HXR LT source,
from its {\sl Yohkoh} soft-X-ray (SXR) images, using filter-ratio
method (from Be119 and Al12 images). This analysis has been done in
the following way: We have integrated SXR fluxes from area $A = 4$
{\sl Yohkoh} pixels [i.e. 4.9 $\times$ 4.9 (arcsec)$^2$] at the
center of HXR LT source. Next we determined diameter, $d$, of the
HXR source according to isocontour $0.5I_{max}$, where $I_{max}$ is
the maximum intensity within the source. We assumed that extension
of the HXR source along the line of sight is also $d$ and calculated
the volume of the emitting plasma as $V = 1.15 A d$ (here 1.15 is a
correction factor which takes into account that actual size of the
HXR source is somewhat larger than that determined from isocontour
$0.5I_{max}$). Mean electron number density was calculated as $N =
\sqrt{EM/V}$ and obtained time-variation of the temperature, $T(t)$,
and density, $N(t)$, are shown in Figures \ref{temp} and \ref{nel}.

Figure \ref{temp} shows that increase of energy release in the LT source began already about 23:10:40 UT. Sharp peaks of temperature about 23:12:30 and 23:16 UT (Figure \ref{temp}) are correlated with peaks in HXRs (Figure \ref{94jan16pgm}b) which confirms that significant part of the energy of accelerated electrons is deposited within the loop-top source (the temperature peaks are somewhat, $\sim$16 s, delayed relative to the HXR peaks which is due to accumulation of plasma heating by accelerated electrons). Significant random fluctuations of $N(t)$ in Figure \ref{nel} are mostly due to random errors in estimates of the source volume from individual HXR images. Systematic increase of density with time seen in Figure \ref{nel} is due to chromospheric evaporation flow.

\begin{figure}
\centerline{\includegraphics[width=1\textwidth]{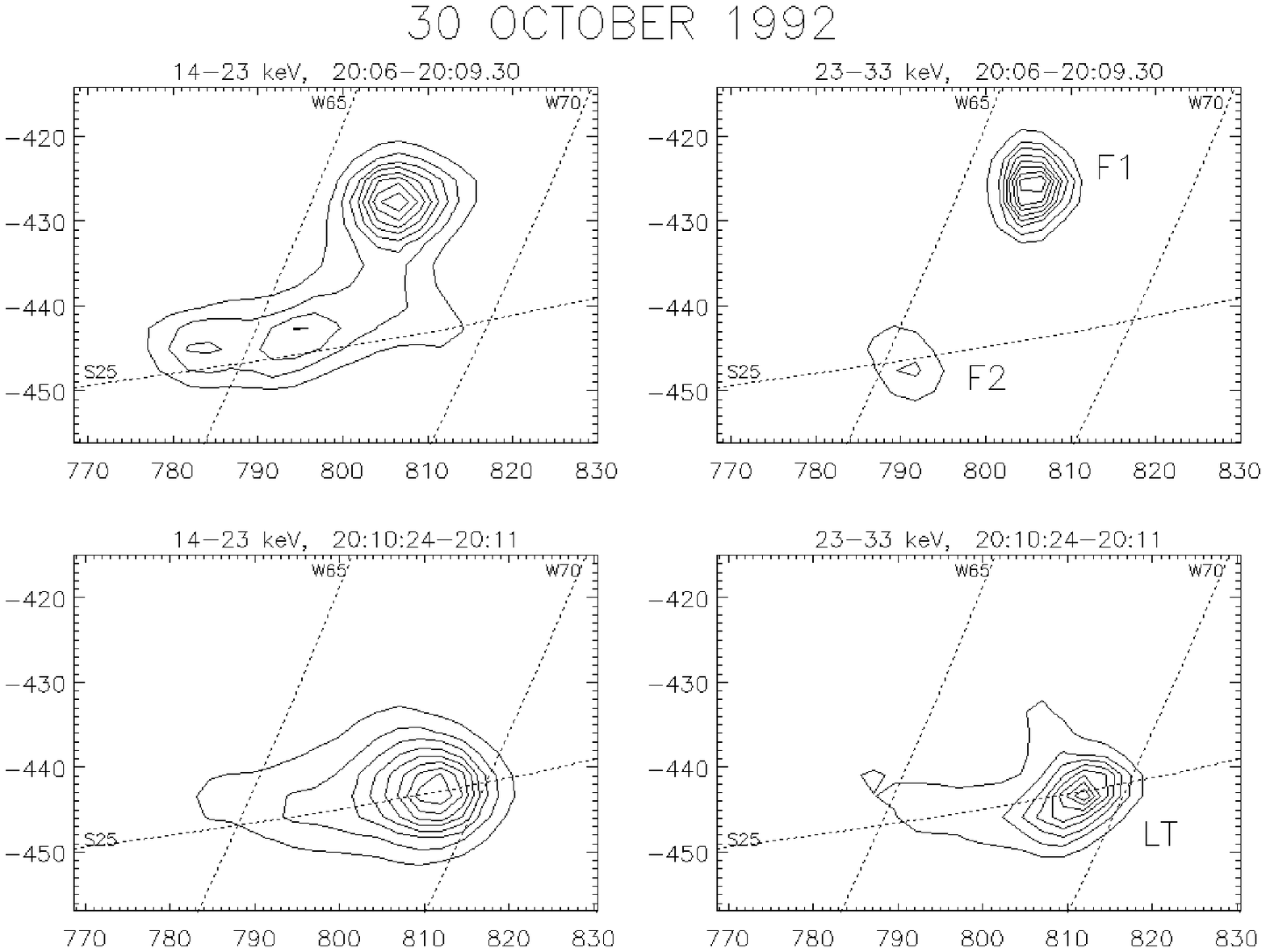}}
\hfill \caption{HXR images of the flare of 30 October 1992 in two
energy channels: 14-23 and 23-33 keV (vertical columns). Upper row:
images recorded before impulsive phase, lower row: images recorded
during impulsive phase.} \label{92oct30img}
\end{figure}

X-ray images for a flare of 30 October 1992 are shown in Figure
\ref{92oct30img}. We see similar behavior like in Figure
\ref{mozaic}: Footpoint sources were stronger than LT source before impulsive phase  and the LT source was stronger during impulsive phase.

\subsection{Determination of magnetic field strength and electron
densities for flares which were investigated in Paper I}
\label{det}

In Paper I we used $N = 1.0 \times 10^{10}$\, cm$^{-3}$ as typical
value of the electron number density in HXR LT sources during
impulsive phase. This value has been obtained by \inlinecite{k+l08}
from {\sl RHESSI} soft-X-ray images of the sources. In Paper I we
used this value of $N$ to estimate magnetic field strength in
oscillating magnetic traps:
\begin{equation}
B_2 = v_2 \sqrt{4 {\pi} {\rho}}
\end{equation}
where $v_2$ is the wave speed estimated from analysis of HXR
oscillations and ${\rho} \approx N m_H$, $m_H$ is the mass of
proton. Obtained values of $B_2$ are given in Table \ref{tab1}.

For eight flares of those investigated in Paper I it was possible to
determine mean electron density, $N$, for HXR LT source from SXR
images using the method described in Section \ref{16jan}. The values
are given in Table \ref{tab1}. We see that most values are
significantly higher than that assumed in Paper I ($N = 1.0 \times
10^{10}$ cm$^{-3}$). This indicates that the value of $N$ used in
Paper I, and therefore also magnetic field strength, $B_2$, were
underestimated. We have calculated corrected values, $B_2^{\ast}$,
of the magnetic field strength using Equation (2) with the new
values of $N$ (see Table \ref{tab1}).

Let us note that values of electron number density, $N$, derived
from {\sl Yohkoh}/ SXT images are reliable, since:
\begin{enumerate}
\item They have been confirmed by independent method (see
\opencite{b+j07}),
\item Determination of $N$ from {\sl Yohkoh}/Al12 images does not
depend significantly on temperature estimates, since instrumental
response function for the Al12 images is nearly constant in wide
range of temperatures (7-40 MK -- see curve {\sl f} in Figure 9 of
\opencite{tsu91}).
\end{enumerate}

\begin{table}
\caption{Values of mean electron density and magnetic field
strength} \label{tab1}
\begin{tabular}{ccccc}
\hline Date & UT & $B_2$ [G]$^{\rm a}$ & $N$ [$10^{10}$ cm$^{-3}$]$^{\rm b}$ & $B_2^{\ast}$ [G]$^{\rm c}$ \\
\hline 28 Jun 92 & 14:00 & 36 & 1.6 & 45 \\
8 May 98 & 01:59 & 40 & 2.3 & 62 \\
31 Oct 91 & 09:11 & 26 & 6.0 & 64 \\
14 May 93 & 22:05 & 30 & 12 & 104 \\
7 Jun 93 & 14:22 & 33 & 15 & 127 \\
16 Jan 94 & 23:17 & 35 & 8.0 & 98 \\
23 Jun 00 & 14:26 & 28 & 15 & 61 \\
25 Nov 00 & 18:39 & 21 & 9 & 63 \\
 \hline
\end{tabular}
\begin{list}{}{}
\item[$^{\rm a}$] magnetic field strength estimated in Paper I
(assuming $N = 1 \times 10^{10}$ cm$^{-3}$)
\item[$^{\rm b}$] mean electron number density in HXR LT source at
HXR maximum, estimated in the present paper
\item[$^{\rm c}$] magnetic field strength calculated with the
new value of $N$
\end{list}
\end{table}

\section{Discussion and summary} \label{sum}

\begin{figure}
\centerline{\includegraphics[width=1\textwidth]{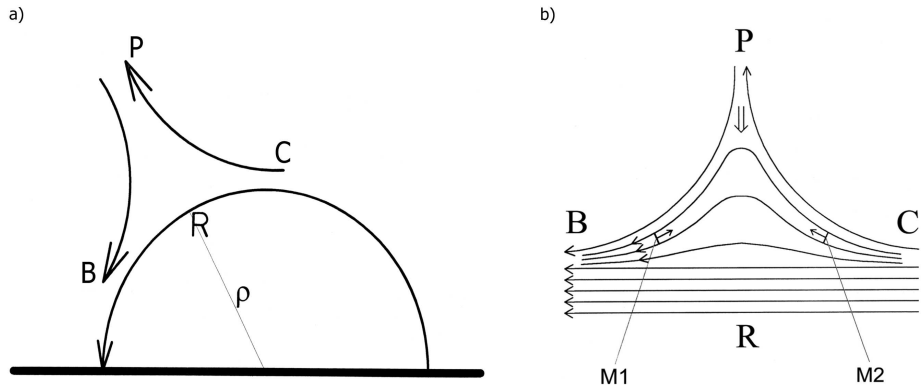}}
\caption{{\bf a} Schematic diagram showing a cusp-like magnetic
configuration. Magnetic fields PB and CP reconnect at P. The thick
horizontal line is the chromosphere. {\bf b} Detailed picture of the
BPC cusp-like magnetic structure.} \label{scheme}
\end{figure}

In Paper I we have proposed a model of oscillating magnetic traps to explain quasi-periodic oscillations seen in HXRs. In this model we have assumed that:
\begin{enumerate}
\item Magnetic structure at the top of flaring loop is triangular (``cusp-like''; Figure \ref{scheme}).
\item Electrons are efficiently accelerated during compression of magnetic traps (\opencite{s+k97},\hspace*{1mm} \opencite{k+k04},\hspace*{1mm} \opencite{b+s05}).
\end{enumerate}

The cusp-like structure is now a standard model of solar flare magnetic field -- see \inlinecite{asc04}. At the top of this structure (near P in Figure \ref{scheme}) magnetic reconnection occurs. Reconnected magnetic field and plasma flow into the confined volume BPC with high velocity. Hence, it seems to be obvious that, this should excite magnetosonic oscillations within the BPC volume.

Magnetic fields PB and PC reconnect at P (Figure \ref{scheme}a). This generates a sequence of magnetic traps which move downward, the traps overtake each other, they collide and undergo compression (Figure \ref{scheme}b). During the compression particles are accelerated within the traps, magnetic pressure, gas pressure, and the pressure of accelerated particles increase, so that the compression is stopped, the traps can expand and undergo magnetosonic oscillations.

 During the compression parameters of the traps undergo strong changes. At the beginning of compression the trap ratio, $\chi = B_{max}/B_{min}$, is high ($B_{max}$ is the magnetic field strength at magnetic mirrors and $B_{min}$ is the strength at the middle of the trap). The trap ratio decreases during compression and it reaches the lowest value, ${\chi}_{min}$, at the end of compression. Then the electrons reach highest energies and they can most easily escape from the trap. Plasma density determines the fraction of accelerated electrons which lose their energy within the trap, i.e. before they escape from the loop-top source. This fraction depends on the energy, since Coulomb losses depend on energy (see changes of the footpoint/loop-top intensity ratio along rows and along columns in Figure \ref{mozaic}). 

Distribution of the oscillations within the BPC volume is determined by the time-profile, $v(t)$, of the velocity of reconnection flow which excites the oscillations. Its central part ($dv/dt \approx 0$) determines a ``main trap'' which contains many electrons and which is responsible for strong HXR pulses. Wings of the velocity profile ($dv/dt \ne 0$) are responsible for excitation of oscillations of many other (``secondary'') traps whose oscillations are shifted in phase. Superposition of many HXR pulses coming from these secondary traps gives a ``quasi-smooth'' emission (seen as the emission below the pulses in Figures \ref{93may14pgm}-\ref{93mar12pgmb}). We see that this quasi-smooth emission is stronger than pulses in the investigated flares which indicates that most of the volume BPC was filled with the ``secondary'' traps (their volume was greater than the volume of main trap).

In many flares HXR pulses are not clearly seen during quick increase of HXR emission, i.e. during beginning of impulsive phase (see example in Figure \ref{93mar12pgmb}). According to our model of oscillating magnetic traps, in such cases the time-profile, $v(t)$ of reconnection flow is quasi-rectangular, i.e. its maximum is flat. Therefore, it excites many oscillating traps of similar power, which have different length (see Figure \ref{scheme}b) and their oscillations are shifted in phase. This gives strong quasi-smooth HXR emission. Clear quasi-periodic pulses are seen only near maximum of HXR emission (see Figure \ref{93mar12pgmb}).

When we observe a long sequence of quasi-periodic oscillations (like in Figures \ref{93may14pgm}-\ref{93mar12pgmb}), this indicates that:
\begin{enumerate}
\item The oscillations have been excited by a short pulse of reconnection flow (otherwise the oscillations would be more chaotic).
\item The oscillations are self-maintained, i.e. some feedback mechanism operates which causes that the oscillations do not decay, but they increase during impulsive phase.
\end{enumerate}

\inlinecite{k+l08} investigated HXR LT sources from {\sl RHESSI} observations and they have found that the energy contained in non-thermal electrons is usually higher than thermal energy. This means that also the pressure of non-thermal electrons is higher than gas pressure, $p_{NT} > p$. This suggests that the pressure, $p_{NT}$, of non-thermal electrons is an important factor in the feedback mechanism which maintains the oscillations of magnetic traps. During compression of a magnetic trap the pressure $p_{NT}$ steeply increases and this causes increase of the amplitude of next expansion of the trap. This, in turn, causes increase of restoring force (i.e. the tension of bent magnetic field lines) and therefore the pulse of the pressure $p_{NT}$ will be stronger during the next compression. Hence, there is a feedback between the intensity of pulses of the pressure $p_{NT}$ and the amplitude of the magnetic trap oscillation and this feedback is responsible for quick increase of the oscillations during impulsive phase.

It may be interesting to note that this mechanism of maintaining the oscillations is analogous to the mechanism of instability which is responsible for stellar pulsations (``non-adiabatic pulsations''): in both cases the point is that additional energy is accumulated during compression of the gas (it is the energy of helium ionization in the case of Cepheids and it is energy of non-thermal electrons in our case).

Main results of the present paper are the following:
\begin{enumerate}
\item It has been confirmed that quasi-periodic oscillations (QPO) occur in HXR emission of solar flares.
\item We have found that low-amplitude QPO occur before impulsive phase of some flares.
\item We have found that quasi-period of the oscillations can change in some flares. We interpret this as being due to changes of the length of oscillating magnetic traps.
\item During impulsive phase most of the energy of accelerated (non-thermal) electrons is deposited within the HXR loop-top sources (Section \ref{16jan}). [For weak HXR flares this is seen at lower energies (14-23 keV), but not at 23-33 keV.]
\item We argue that the basic properties of the HXR oscillations can be explained in terms of a simple model of oscillating magnetic traps (see Paper I). This model allows us also to explain large number of electrons which are accelerated during impulsive phase: Observations show that the amplitude of the oscillations quickly increases and therefore the traps are filled with increasing amount of plasma coming from chromospheric evaporation. This is main source of electrons which undergo acceleration.
\item We suggest that a feedback between the pressure of accelerated electrons and the amplitude of the following expansion of magnetic trap is the mechanism which causes the quick increase of the amplitude of oscillations.
\item We have also determined improved values of electron number density and magnetic field strength for HXR loop-top sources of several flares which were investigated in Paper I. Obtained values fall within the limits of $N \approx (2 -15) \times 10^{10}$ cm$^{-3}$, $B \approx (45 - 130)$ gauss.
\end{enumerate}

Main advantages of the model of oscillating magnetic traps are the following:
\begin{enumerate}
\item Acceleration of electrons occurs in a large volume (BPC in Figure \ref{scheme}).
\item During development of impulsive phase plasma is delivered into BPC volume by chromospheric evaporation flow whose density is higher than the density of surrounding corona. This explains why number of accelerated electrons is very high.
\end{enumerate}

\begin{acks}
The {\sl Yohkoh} satellite is a project of the Institute of Space and Astronautical Science of Japan. The {\sl Compton Gamma Ray Observatory} is a project of NASA. We would like to thank anonymous referee for valuable remarks which helped us to improve this paper. This work was supported by Polish Ministry of Science and High Education grant No. N\,N203\,1937 33.
\end{acks}

\end{article}

\end{document}